\documentclass[nonacm,sigconf,10pt]{acmart}

\renewcommand\footnotetextcopyrightpermission[1]{} 
\AtBeginDocument{%
  \providecommand\BibTeX{{%
    \normalfont B\kern-0.5em{\scshape i\kern-0.25em b}\kern-0.8em\TeX}}}

\setcopyright{none}

\newcommand{\sysname}{DIT}

\usepackage{subfig}
\usepackage{gensymb}

\usepackage{makecell}

\usepackage{tikz}  
\usetikzlibrary{spy}
\usepackage{color, colortbl}
\usepackage{pgfplots}
\pgfplotsset{compat=newest}
\usepackage{multicol}
\usepackage{circuitikz}
\usepackage[nolist]{acronym}
\usepgfplotslibrary{groupplots}
\usetikzlibrary{shapes}
\usepackage{siunitx}  
\usepackage{graphicx}

\definecolor{mittelblau}{RGB}{0, 126, 198}
\definecolor{violettblau}{cmyk}{0.9, 0.6, 0, 0}
\definecolor{rot}{RGB}{238, 28 35}
\definecolor{apfelgruen}{RGB}{140, 198, 62}
\definecolor{gelb}{RGB}{1, 221, 0}
\definecolor{orange}{RGB}{244, 111, 33}
\definecolor{pink}{RGB}{237, 0, 140}
\definecolor{lila}{RGB}{128, 10, 145}
\definecolor{hellgrau}{RGB}{224, 224, 224}
\definecolor{mittelgrau}{RGB}{128, 128, 128}
\definecolor{dunkelgrau}{RGB}{80,80,80}
\definecolor{anthrazit}{RGB}{19, 31, 31}
\definecolor{darkgreen}{RGB}{0.125,0.5,0.169}

\usepackage{ifthen}
\usetikzlibrary{matrix,calc}

\usetikzlibrary{external}
\usepackage[colorinlistoftodos]{todonotes}

\usepackage[ruled]{algorithm2e} 

\usetikzlibrary{external}
\begin{document}
\begin{multicols}{2}
\begin{acronym}[WSSUS]

    \acro{5G}{fifth-generation}
    \acro{ADC}{analog to digital converter}
    \acro{AFE}{analog front end}
    \acro{AGC}{automatic gain control}
    \acro{AGV}{automated guided vehicle}
    \acro{AMP}{approximate message passing}
    \acro{AWGN}{additive white Gaussian noise}

    \acro{BER}{bit error rate}
    \acro{BB}{baseband}
    \acro{bpcu}{bits per channel use}
    \acro{BP}{belief propagation}
    \acro{BPSK}{binary phase shift keying}
    \acro{BS}{base station}

    \acro{CB}{codebook}
    \acro{CDF}{cumulative distribution function}
    \acro{CFO}{carrier frequency offset}
    \acro{CoSaMP}{compressive sampling matching pursuit}
    \acro{CP}{cyclic prefix}
    \acro{CS}{compressive sensing} 
    \acro{CSI}{channel state information}
    \acro{CNN}{convolutional neural network}

    \acro{DA}{domain adaptation}
    \acro{DAC}{digital-analog-converter}
    \acro{DC}{direct current}
    \acro{DE}{distance error}
    \acro{DeepL}{deep-learning}
    \acro{DoF}{degree-of-freedom}
    \acro{DFT}{discrete Fourier transformation}
    \acro{DL}{deep learning}
    \acro{DS}{delay spread}
    \acro{DSP}{digital signal processing}

    \acro{ECC}{error-correcting code}
    \acro{ENoB}{effective number of bits}
    \acro{ERP}{effective radiated power}
    \acro{EVM}{error vector magnitude}

    \acro{FB}{feedback}
    \acro{FC}{fully connected}
    \acro{FDD}{frequency division duplexing}
    \acro{FDM}{frequency division multiplexing}
    \acro{FIR}{finite impulse response}
    \acro{FT}{fine tuning}
    \acro{FPGA}{field programmable gate array}
    \acro{GAN}{Generative adversarial network}
    \acro{GPIO}{general-purpose input/output}
    \acro{GPS}{global positioning system}
    \acro{GPSDO}{GPS disciplined oscillator}
    \acro{GPU}{graphical processing unit}

    \acro{HDD}{hard decision decoding}
    \acro{IC}{integrated circuit}
    \acro{I2C}{Inter-Integrated Circuit}
    \acro{ICSP}{in-circuit serial programming}
    \acro{IF}{intermediate frequency}
    \acro{i.i.d.}{independent and identically distributed}
    \acro{IIR}{infinite impulse response}
    \acro{IMU}{inertial measurement unit}
    \acro{IoT}{Internet of Things}
    \acro{IPS}{indoor positioning system}
    \acro{IR}{infrared}
    \acro{JSDM}{Joint Spatial Division and Multiplexing}
    \acro{LLR}{log-likelihood ratio}
    \acro{LP}{leakage precoder}
    \acro{LMMSE}{Linear Minimum Mean Square Error}
    \acro{LO}{local oscillator}
    \acro{LoS}{line of sight}
    \acro{LiDaR}{Light Detection and Ranging}
    \acro{LS}{least squares}
    \acro{LSTM}{long-term short-term memory}
    \acro{LTE}{Long Term Evolution}
    \acro{LTI}{linear time invariant}
    \acro{LTV}{linear time variant}
  
    \acro{MAP}{maximum a posteriori}
    \acro{MDE}{mean distance error}
    \acro{MDA}{mean distance accuracy}
    \acro{MEMS}{Micro-Electro-Mechanical Systems}
    \acro{MIMO}{multiple input multiple output}
    \acro{MISO}{multiple input single output}
    \acro{ML}{maximum likelihood}
    \acro{MLD}{maximum likelihood decoding}
    \acro{mMIMO}{massive multiple input multiple output}
    \acro{MMSE}{minimum mean square error}
    \acro{M-MMSE}{multi-cell minimum mean square error}
    \acro{MR}{maximum ratio}
    \acro{MRC}{maximum ratio combining}
    \acro{MRP}{maximum ratio precoding}
    \acro{MRT}{maximum ratio transmission}
    \acro{MSE}{mean squared error}
    \acro{MQTT}{Message Queuing Telemetry Transport}
    \acro{MU}{multi-user}
    \acro{NF}{noise figure}
    \acro{NN}{Neural Network}
    \acro{NNI}{Neural Network Intelligence}
    \acro{NLoS}{non-line of sight}
    \acro{NND}{neural network decoding}
    \acro{NTP}{Network Time Protocol}
    \acro{NMSE}{normalized mean squared error}
    \acro{NU}{not-used}
    \acro{OFDM}{orthogonal frequency division multiplex}
    \acro{OMP}{orthogonal matching pursuit}
    \acro{OPS}{outdoor positioning system}
    \acro{OT}{optimal transport}
    \acro{PB}{passband}
    \acro{PCB}{printed circuit board}
    \acro{PDR}{pedestrian dead reckoning}
    \acro{PDF}{probability density function}
    \acro{PDP}{power-delay-profile}
    \acro{PLL}{phase-locked-loop}
    \acro{PO}{phase-only}
    \acro{PPS}{pulse per second}
    \acro{QPSK}{quadrature phase shift keying}
    \acro{QuaDRIGa}{Quasi Deterministic Radio Channel Generator}

    \acro{ReLU}{rectified linear unit}
    \acro{RF}{radio frequency}
    \acro{RMS-DS}{Root Mean Square - Delay Spread}
    \acro{RNN}{recurrent neuronal network}
    \acro{RSSI}{received signal strength indicator}
    \acro{R-ZF}{regularized zero-forcing}
    \acro{SDD}{soft decision decoding}
    \acro{SDR}{software defined radio}
    \acro{SE}{spectral efficiency}
    \acro{SFO}{sampling frequency offset}
    \acro{SLAM}{Simultaneous Localization and Mapping}
    \acro{SGD}{stochastic gradient descent}
    \acro{SISO}{single input single output}
    \acro{SINR}{signal-to-interference-and-noise-ratio}
    \acro{SIR}{signal-to-interference-ratio}
    \acro{SLNR}{signal-to-leakage-and-noise ratio}
    \acro{SNR}{signal-to-noise-ratio}
    \acro{SP}{subspace}
    \acro{SQR}{signal-to-quantization-noise-ratio}
    \acro{SQNR}{signal-to-quantization-noise-ratio}
    \acro{SVD}{singular value decomposition}
    \acro{SU}{single-user}
    \acro{TDD}{time division duplexing}
    \acro{TRIPS}{time-reversal IPS}
    \acro{UE}{user equipment}
    \acro{UL}{uplink}
    \acro{ULA}{uniform line array}
    \acro{URLLC}{ultra-reliable low-latency communication}
    \acro{US}{uncorrelated scattering}
    \acro{USRP}{universal software radio peripheral}
    \acro{UWB}{ultra-wideband}
    \acro{WiFi}{Wireless Fidelity}
    \acro{WSS}{wide sense stationary}
    \acro{WSSUS}{wide sense stationary uncorrelated scattering}

    \acro{ZF}{zero forcing}
\end{acronym}
\end{multicols}

\title{Indoor positioning systems: Smart fusion of a variety of sensor readings}


\author{Maximilian Arnold}
\affiliation{%
    \institution{Nokia Bell Labs \country{Germany}}
    \streetaddress{1 Th{\o}rv{\"a}ld Circle}
    \city{Stuttgart}
    \country{Germany}
}
\email{maximilian.arnold@nokia-bell-labs.com}

\author{Frank Schaich}
\affiliation{%
    \institution{Nokia Bell Labs \country{Germany}}
    \streetaddress{1 Th{\o}rv{\"a}ld Circle}
    \city{Stuttgart}
    \country{Germany}
}
\email{frank.schaich@nokia-bell-labs.com}

\begin{abstract}
Robust and versatile localization techniques are key to the success of the next industrial revolution. Yet, it is uncertain which combination of sensors will be the most robust and valuable. Thus, we present a versatile and reproducible measurement system incorporating a manifold number of state-of-the art sensors to compare and fuse the raw input data. It is shown that some techniques show very good results on the same scenario and data-set, but fall apart on translating to a slightly different scenario. In general we show that the vanilla approach to fuse the raw data achieves reasonable results in the generalization domain, demonstrating that \ac{RF} localization techniques in combination with an \ac{IMU} could result in a very robust and promising candidate for solving this challenging task.
\end{abstract}

\maketitle

\section{Introduction}
Robust and versatile \acp{IPS} are key to the success of the next industrial revolution \cite{Ayyala20_DLoc,Brossard20_AI-IMU}. It can be seen as an key enabler for a wide range of applications such as indoor navigation, smart factories, or it could even provide a basic security functionality in distributed \ac{IoT} sensor networks \cite{chen2017cmIPS,chen2017ConFI,LeeandHahn2017}. In general \acp{IPS}  will incorporate a manifold of different sensors, to enhance the different systems and creating a practical system. Although \acp{IPS} are currently holding back fully autonomous factories due to their reliability regarding their accuracy, it is uncertain which combination of sensors will be the most efficient in terms of cost and usability.

Different robust \ac{IPS} techniques have emerged ranging from trilateration \cite{Hoffmann2020} over angle-of-arrival \cite{wu2015TR} to recent \ac{DL} approaches \cite{8661318}. All of them demonstrate promising results in large parts of a given area, but suffer in specific parts of the environment. Therefore approaches to fine-tune \acp{NN} \cite{9048863} or advanced fusion techniques \cite{FuseOverview,weber2020neural} were proposed. 

Yet, to the best of our knowledge we are the first to present a data-set containing a manifold of off-the-shelf \ac{RF}, magnetic and \ac{IMU} localization techniques being compared in a fair, and neutral manner allowing us to analyze and rate the different techniques. Thus, this work emphasises on \emph{how} to automatically generate a large data-set in different scenarios and compare various approaches for fusion. We demonstrate the feasibility of synchronizing the data over \ac{NTP} and \ac{MQTT} protocols and verify the sanity of our measurement data using classical localization techniques. 
Moving towards robust positioning systems a vanilla state-of-the-art \ac{DL} approach on each of the modalities demonstrate the applicability of these systems. Further we consider a raw-data fusion method using \ac{DL} for enhancing the overall localization performance in all scenarios. 

Due to their underlying structure, \ac{DL} approaches tend to focus heavily on a specific scenario, losing generality and the ability to be transferred to other domains. Thus, we measure two seperate data-sets within one day, with a slightly different setup and try to predict from one to the other. We demonstrate that this is in general possible but with a certain amount of loss. 

To sum up this paper main contributions are:
\begin{itemize}
\item presenting a method to create automatically a large amount of labeled data in a plug and play fashion, extending to a manifold of sensors
\item a data-set containing synchronized sensor readings for various techniques \ac{UWB}-range, \ac{CSI}, \ac{RSSI}, magnetic information, Odometry data
\item \ac{IMU} is shown to be a very valuable element for raw-data fusion, as its independent of the environment 
\item \ac{CSI} and \ac{IMU} data fused together show the best generalisability and enabling efficient \ac{IPS}.
\end{itemize}

\section{Positioning systems}
Tackling the challenging task of indoor localization a manifold of techniques have been proposed, built and tested, yet a clear vision, as e.g. available in outdoor scenarios like the \ac{GPS}, is missing. These proposed techniques can be classified into three classes: (1) \ac{RF}-based localization techniques, e.g. \ac{CSI} or \ac{UWB}; (2) Inertial tracking, e.g. exploiting inertial sensors and (3) camera/infrared/laser based techniques, e.g. \ac{LiDaR}.

Resulting from the fact that \ac{IoT} devices demand battery lives over multiple months/years, these techniques have to be lightweight or computational heavy algorithms have to be shifted out, e.g. to an edge cloud, allowing for more complex and diverse techniques, ranging from trilateration over fingerprinting techniques to advanced \ac{NN} structures. 
 
However, all of these techniques suffer from different deficiencies; \ac{RF} based techniques can experience outages, while inertial tracking drifts over time and camera/\ac{LiDaR} systems are expensive and suffer under low lighting conditions and blockage. 
 Thus, in general a robust \ac{SLAM} algorithm using multiple sensors to create precise maps is missing. To tackle this challenge multiple fusion approaches were envisioned ranging from Kalman filter over particle filter to conditional random fields, all allowing to fuse the data in a raw or processed format. As localization algorithms typically are hand-crafted towards their application, the common method is to fuse the data not in the raw but in a pre-processed fashion. We consider here the typical, yet broadly used, classical techniques and introduce a straightforward and simple fusion method at the raw data level. 
 
\textit{Trilateration}\\
Trilateration was one of the first localization techniques allowing to pinpoint a user in a mobile network with a coarse accuracy. This technique is commonly used in \ac{GPS} devices and has been heavily improved over time. 

\begin{figure}[t]
    \centering
    \includegraphics[]{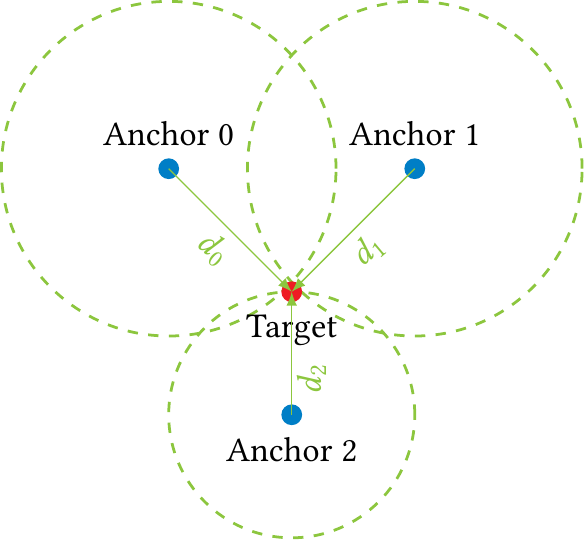}
    \caption{Trilateration example}
    \label{fig:tirlat}
\end{figure}

Fig.~\ref{fig:tirlat} depicts the basic concept of trilateration. Hereby the three distances $d_0,d_1,d_2$ are estimated through measuring time of flights between the target and the different anchors.  The estimated distances can be represented as circles around  the anchors, which intersect in the best case in only a single point, the target location $x,y$. For a more simplified solution the anchor 0 is used as reference point, e.g. $(x_0,y_0)$ is set to 0 by shifting the remaining coordinates by the point, resulting in
\begin{align*}
d_0^2 &= x_\text{target}^2 + y_\text{target}^2\\
d_1^2 &= (x_\text{target}-x_1-x_0)^2 + (y_\text{target}-y_1-y_0)^2 \\
d_2^2 &= (x_\text{target}-x_2-x_0)^2 + (y_\text{target}-y_2-y_0)^2 \\
\end{align*}
As the distance measurements are noisy the circles usually are not intersecting in a single point and thus a different approach needs to be applied, by first intersecting two circles; forming a line, which can be intersected with the remaining circle. This, will execute trilateration well, even if the circles do not intersect perfectly.
\textit{Fingerprinting:}\\
Fingerprinting is a method to localize a target in a well defined-environment, where a map is created in an initial training phase to be used in the deployment phase to locate the target.
Thus, a cumbersome training phase creating an $N$ dimensional vector per position is key for the success of this technique. In the deployment phase the measured $N$ dimensional vector is correlated to the data set. Then either the closest trained vector or an interpolation between the closest trained vectors is used to estimat the position. Due to the huge amount of overhead, this technique is slow, very susceptible to changes of the environment and not considered in most cases.

\textit{Deep Learning positioning:}\\
\ac{DL} techniques are getting more and more applied in many different applications ranging from image classification to speech detection. Its main feature is to map an input space to an output space over an regression or classification task, where classical models are either too complex, too susceptible for different impairments or too computational heavy. 
Thus in this case a non-linear transformation 
\begin{align}
    \bf{m} \xrightarrow{f_{_\Theta}} (\hat{x},\hat{y}) \notag,
\end{align}
has to be learned, which maps the input measurement vector $\bf{m}$ to a discrete position, depending on the hyperparameters $\Theta$. Thus allowing for a system learning the environment and the underlying structure of the measurements. 
\begin{figure}[H]
    \centering
    \includegraphics[]{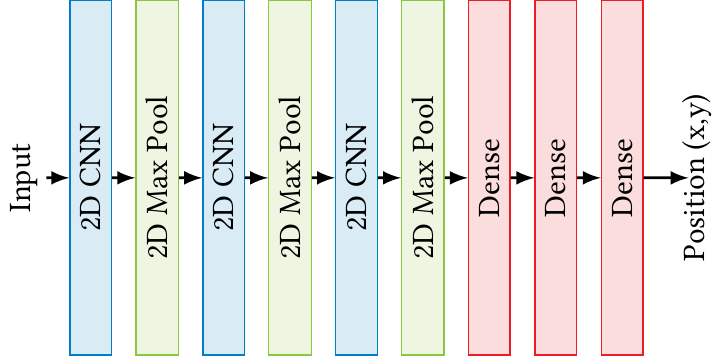}
    \vspace*{-0.4cm}
    \caption{\ac{NN} architecture optimized and searched via AutoML tools.} 
    \label{fig:DLArchitecture}
\end{figure}
Fig.~\ref{fig:DLArchitecture} depicts the \ac{NN} architecture searched via the auto machine-learning (AutoML) tool from Microsoft \cite{NNI}. The hyperparameters were also optimized via this tool, allowing for a defined comparison for each technique\footnote{The hyper-parameters and layer sizes will be distributed with the dataset}. This \ac{NN} will be used throughout the remaining paper with optimized kernel/pool/stride/dense sizes based on the input measurements using the tool \cite{NNI}.

\section{Measurement System} \label{sec:meas_sys}
The success for \ac{DL} in vision and audio based techniques was achieved only by leveraging the large datasets being already available for classical algorithms. For creating a highly used data-set for \ac{IPS}, we introduce a versatile measurement setup, allowing us to add in a plug-and-play fashion any kind of sensor.
\begin{figure}[H]
    \centering
    \includegraphics[]{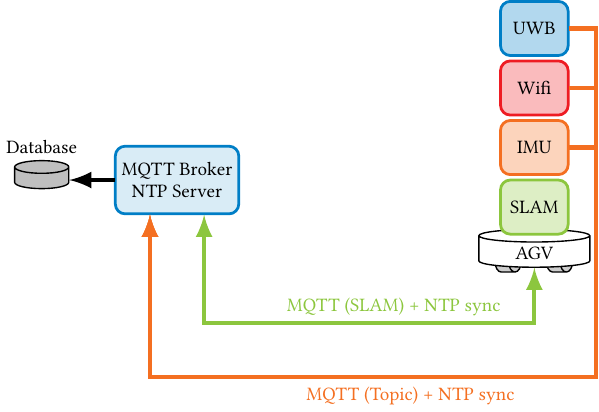}
    \vspace*{-0.4cm}
    \caption{Versatile measurement system.} 
    \label{fig:MeasurementSystem}
\end{figure}
Fig.~\ref{fig:MeasurementSystem} depicts the flexible and versatile measurement platform for automatically creating a large amount of labelled data. It is based on a vacuum cleaner robot including a two-dimensional \ac{LiDaR}, ultrasonic distance sensors, wheel-tick sensors, multiple \ac{IR} sensors, and an \ac{IMU}. 
These modalities are fused together---via a Kalman filter-based \ac{SLAM} algorithm---to generate pseudo-groundtruth for our subsequent experiments.
The \ac{SLAM} accuracy is better than \SI{1}{\centi\metre}, as per the evaluation conducted by Hoffmann et al.~\cite{Hoffmann2020}.
This level of accuracy satisfies the requirements of most localization applications. 

In general we are creating the system by using: (1) an robot for the pseudo-groundtruth, (2) a platform  mounted on top of the robot to host additional sensors, (3) a \ac{MQTT} broker with a database for data acquisition, and (4) an \ac{NTP} server for time synchronisation between various sensors.
In terms of hardware, the platform was fabricated inhouse from 3D printed parts and a LEGO building plate. 
On the software side, a \ac{MQTT} broker and a \ac{NTP} server form the core of the data acquisition system. 
Specifically, each sensor synchronises its time-clock with the local \ac{NTP} server, which allows for an accuracy below \SI{1}{\milli\second}~\cite{Milis1991}. 
The synchronised measurements are then streamed to the \ac{MQTT} broker. 
By virtue of \ac{MQTT} and the back-end database, our sensor system is modular and extensible.

\begin{figure}[H]
    \centering
    \includegraphics[]{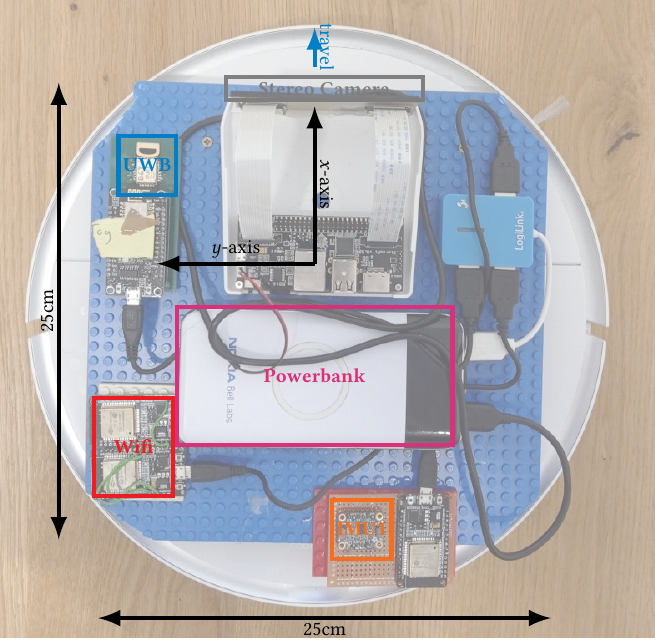}
    \vspace*{-0.4cm}
    \caption{Build up of measurement platform consisting of: A roborock vacuum cleaner; a building platform; a modality of sensors and a power-bank.} 
    \label{fig:BuildupRoborock}
\end{figure}
We leverage a Roborock S50 from Xiaomi as base robot platform shown in the build-up figure \ref{fig:BuildupRoborock}. The robotic platform streams its pseudo-ground-truth positions to the \ac{MQTT} broker with a \SI{5}{\hertz} update rate. A power-bank of 10Ah is used to power the sensors. We utilise \ac{IMU} modules from Adafruit~\cite{AdaFruit} equipped with Bosch BNO055 inertial chipsets~\cite{IMUManual}. 
Each \ac{IMU} module is connected to an ESP32 \SI{2.4}{\giga\hertz} WiFi chip-set over an I2C interface. The \ac{UWB} system consists of three anchors and one tag using the DW1000 chip-set. All of them are using the band 5 with \SI{500}{\mega \hertz} bandwidth. The WiFi CSI and RSSI is achieved by using two ESP32 listening to the pilot WiFi signals of the surrounding anchors. The anchors are all ESP32 creating traffic by constantly pinging the Fritzbox router.

\subsection{Pseudo groundtruth}\label{Sec:MeasSysGT}
\begin{figure}[t]
    \centering
    \includegraphics[]{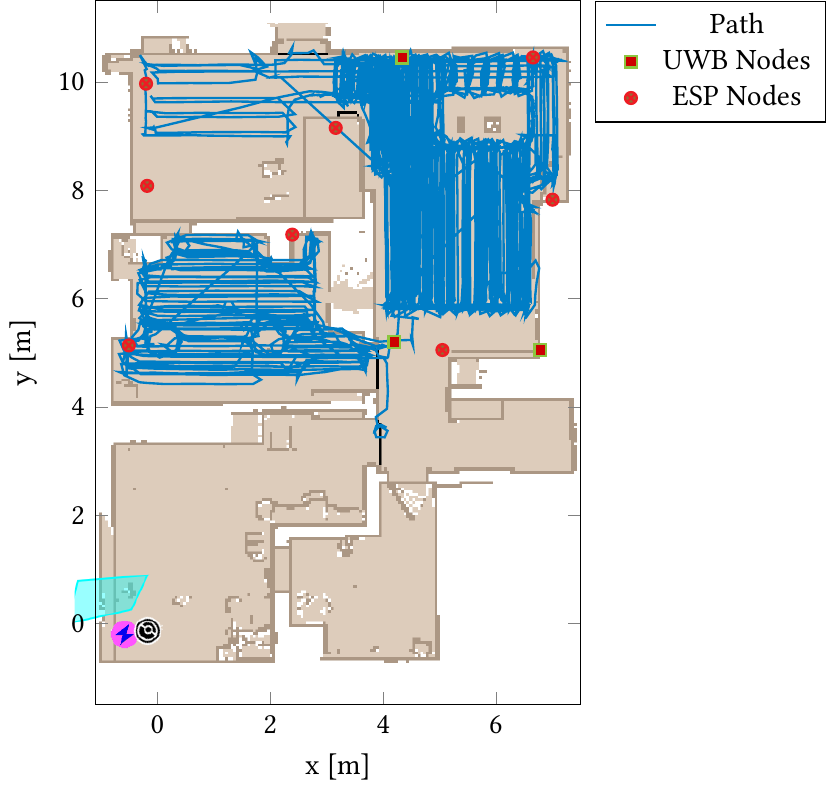}
    \vspace*{-0.4cm}
    \caption{Measurement paths and anchor placements in the flat.} 
    \label{fig:MeasPath}
    \vspace{-0.4cm}
\end{figure}
Fig.~\ref{fig:MeasPath} depicts the measurement area within a flat's floor-plan located in Stuttgart, Germany. 
The robot trajectories are highlighted in blue. 
The floor-plan has been automatically created by the robotic \ac{SLAM} algorithm as described above.
The measurement campaign was conducted over a single day creating one data-set in the morning and another data-set in the evening.
Coordinate translation of the \ac{SLAM} pseudo ground-truth is needed in order to compensate for all sensor positions, as the robot can spin around at a certain location moving the sensors on a circular trajectory.
In order to correct for these effects, we devised the following. Firstly, the coordinate system of the \ac{SLAM} calculates the position based on an anchor point located at the front of the robot. The anchor point is translated to the geometrical center of the robot.
The \ac{SLAM} ground-truth coordinates $(x_{_\text{SLAM}}, y_{_\text{SLAM}})$ and orientation $\phi_{_\text{SLAM}}$ are referenced to the robot coordinate system. Denoted by $(\hat{x}_{_\text{SLAM}}, \hat{y}_{_\text{SLAM}})$ and  $\hat{\phi}_{_\text{SLAM}}$ we get a translated reference coordinate and orientation system, respectively. 
Then the offset angle $\phi_{_{\text{Sensor},\ell}}$ and coordinates $(x_{_{\text{Sensor},\ell}}, y_{_{\text{Sensor},\ell}})$ of the sensor  were measured.
Finally, the measured values are used to translate the position of the sensor to the global coordination system, according to
\begin{align}
\label{eqn:Sensor Positions}
\begin{split}
r_{_{\text{Sensor},\ell}} &=  \sqrt{x^2_{_{\text{Sensor}, \ell}} +  y^2_{_{\text{Sensor},\ell}}}
\\
 \hat{x}_{_{\text{Sensor},\ell}} &= \hat{x}_{_\text{SLAM}} + \text{Re}\Bigl\{ r_{_{\text{Sensor},\ell}} \; e^{j\left(\hat{\phi}_{_\text{SLAM}}+\phi_{_{\text{Sensor},\ell}}\right)} \Bigr\}
\\
 \hat{y}_{_{\text{Sensor},\ell}} &= \hat{y}_{_\text{SLAM}} + \text{Im}\Bigl\{ r_{_{\text{Sensor},\ell}} \; e^{j\left(\hat{\phi}_{_\text{SLAM}}+\phi_{_{\text{Sensor},\ell}}\right)} \Bigr\}
\end{split}
\end{align}

\begin{table}[t]
\centering
\begin{tabular}{|c|c|c|}
\hline 
Parameter       & Dataset 1 & Dataset 2 \\ \hline \hline
Time Run        & \SI{160.17}{\min}            &  \SI{141.0}{\min}                          \\ \hline
Path Covered    & \SI{1975.28}{\metre}         & \SI{1733}{\metre}                           \\ \hline
Data            & Morning                        & Afternoon                       \\ \hline
Nb. CSI Anchor  & 13 (ESP)           & 13 (ESP)          \\ \hline
Nb. UWB Anchor  & 3 (DW1000)                    & 3 (DW1000)                     \\ \hline
Nb. IMU         & 1 (Bosch BNO05)              & 1 (Bosch BNO05)                \\ \hline
Nb. Cameras     & 0                              & 1 (Pi Stereo Hat)              \\ \hline
Labels          & 2D \ac{LiDaR}                       & 2D \ac{LiDaR}                       \\ \hline
Update Rate CSI & \SI{7.50}{\hertz}                           & \SI{8.5}{\hertz}                            \\ \hline
Update Rate UWB & \SI{9}{\hertz}              & \SI{7}{\hertz}                              \\ \hline
Update Rate IMU & \SI{76.93}{\hertz}              & \SI{76.93}{\hertz}                          \\ \hline
Nb points CSI   & 34417                          & 46584                          \\ \hline
Nb. points UWB  & 52152                          & 28239                          \\ \hline
Nb. points IMU  & 747853                         & 656640                        \\ \hline
Data ESP  &  CSI + RSSI   &  CSI + RSSI                        \\ \hline
Data UWB   & Range + Power                          & Range + Power                          \\ \hline
Data IMU  & 9-DoF  +Odo                       & 9-DoF  +Odo                        \\ \hline
\end{tabular}
\caption{Two measured datasets used for localization fusion.}
\label{table:LoCDatasets}
\vspace{-0.8cm}
\end{table}
Tab.~\ref{table:LoCDatasets} depicts the two available measurement data-sets used for analyzing sensor fusion and positioning algorithms.We have repeated the same measurement twice to understand if the algorithms  generalize or not. Thus, a system trained on Dataset 1 should be able to infer the position on the Dataset 2. The difference between the two data-sets is is not exactly the same measurement through slightly changing the building plate while adding additional sensors. This will give a strong indication on the robustness and versatility of the system. 

\begin{figure}[H]
    \centering
    \includegraphics[]{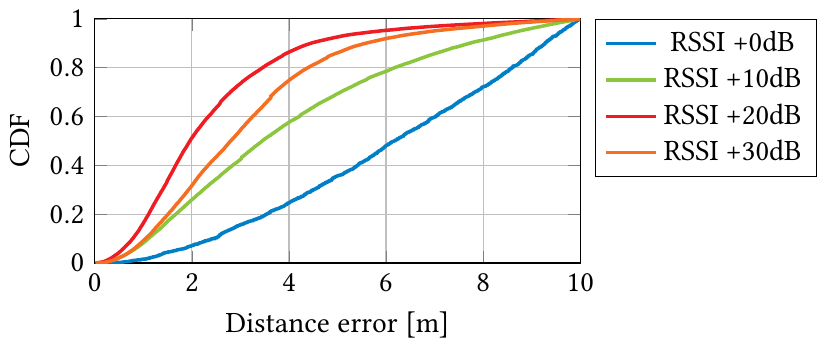}
    \vspace*{-0.4cm}
    \caption{Importance of RSSI calibration for trilateration based on distance.} 
    \label{fig:RSSICalibration}
\end{figure}
Emphasizing that this is raw data directly streamed from the sensor, we highlight the necessity of calibration and fusing the sensors correctly, like in an actual deployment (comp. Fig. \ref{fig:RSSICalibration}). Hereby the \ac{RSSI} per node was used to estimate the distance to the target and the three closest \ac{RSSI} sources were used to do trilateration. It turns out that the \ac{RSSI} in this setup performs the best if every \ac{RSSI} measurement is enhanced by 20dB allowing for a more robust trilateration.

\section{Results} \label{sec:evaluation}
A practical localization system in general requires to achieve: (1) at least in 99\% an distance error to be smaller than \SI{1}{\metre}; (2) being able to handle small changes in the environment, e.g. translating from one data-set to another. First, the general performance of the single algorithms is evaluated; then we investigate sensor fusion and finally the we analyze the generalizability over time. 
\begin{figure}[H]
    \centering
    \includegraphics[]{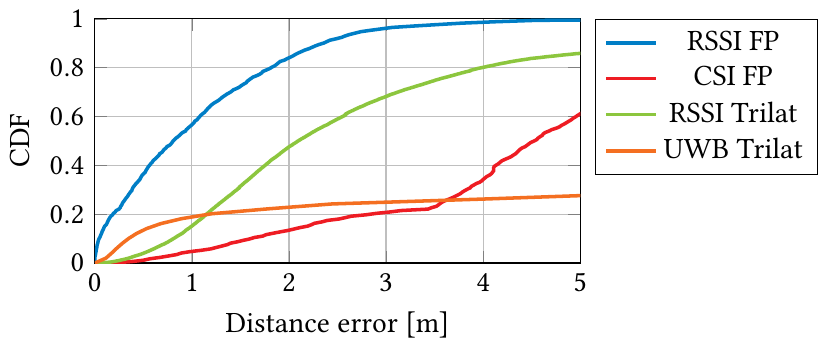}
    \vspace*{-0.4cm}
    \caption{Performance of classical methods on the Dataset 1.} 
    \label{fig:ClassicalMethods}
    \vspace{-0.4cm}
\end{figure}
Fig.~\ref{fig:ClassicalMethods} the \ac{CDF} of the distance error for several variants of a fingerprinting-based (FP) method and two trilateration methods. It becomes apparent that due to the limited number of \ac{UWB} anchors the trilateration succeeds only in 20 \% of the cases, indicating there is an area with a higher accuracy while the remaining area is only poorly covered by three nodes. The RSSI trilateration using the virtue of more nodes achieves a better accuracy, yet still only in a similar range as commercial \ac{GPS} devices. The \ac{CSI} fingerprinting methods fails, as the phase of the anchors is not fixed and thus an arbitrary phase shift between nodes is created. An advanced classical methods might be able to track this, but this would exceed the scope of this paper. The RSSI FP method gives a good indication what can be achieved with a higher number of nodes in the \ac{UWB} system. This indicates a high likelihood for fusion of methods to join the advantages of these methods and remove unambiguous features. 
\begin{figure}[H]
    \centering
    \includegraphics[]{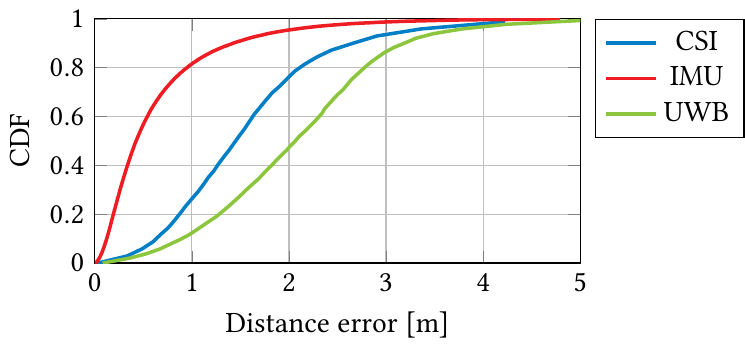}
    \vspace*{-0.4cm}
    \caption{Performance of vanilla \ac{NN} on the different raw inputs from Dataset 1.} 
    \label{fig:NNTechniques}
    \vspace{-0.4cm}
\end{figure}
Fig.~\ref{fig:NNTechniques} depicts the results of the vanilla \ac{NN} optimised using \ac{NNI}, where the \ac{CSI}, \ac{IMU} (3D magnetic, 3D acceleration, 3D gyro) and \ac{UWB} (range) data is feed into the \ac{NN} and optimized using a 90/10 split between training and testing. It becomes apparent that the \ac{UWB} performance is better due to the ability to filter out points with only one connect node allowing to use the middle as the estimate position, thus limiting the error. Moreover the \ac{CSI} method seems to be achieving better performance than its classical pendant the fingerprinting method, due to more information being available and finding a common function for the features. Due to the magnetic fingerprinting, the performance of the vanilla \ac{IMU} \ac{NN} seems to be very solid. Note that tracking in this case is not possible, as the data is shuffled. The tracking ability will be used in the last section, by slicing a certain amount of time together. In general the techniques achieve similar performances as the classical methods but removing the outliers.  

\subsection{Fusion}
To remove drift and to enhance performance the raw data of the various techniques is stacked together and the effective performance of a vanilla \ac{NN} is estimated.
\begin{figure}[H]
    \centering
    \includegraphics[]{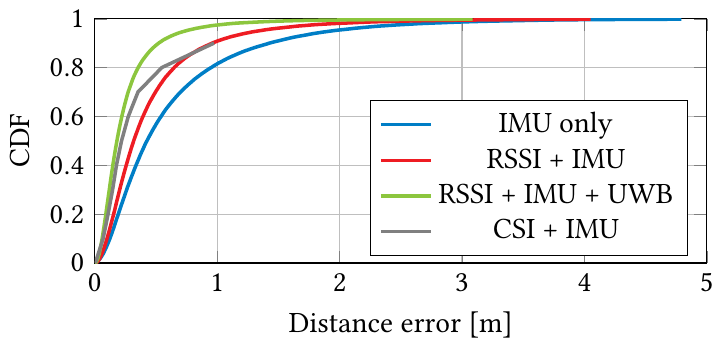}
    \vspace*{-0.4cm}
    \caption{Performance of the vanilla \ac{NN} on the stacked raw data.} 
    \label{fig:FusedVanillaNN}
    \vspace{-0.4cm}
\end{figure}
Fig.~\ref{fig:FusedVanillaNN} includes the \ac{IMU} curve as  reference (blue). Each addition of further sensor data improves the system performance as expected. In general the \ac{RSSI} in combination with the \ac{IMU} data shows potential in removing the outliers in the \ac{IMU} system. The CSI in combination with the \ac{IMU} shows also reasonable results in average below than \SI{10}{\centi\metre} and in 95\% better than \SI{1}{\metre} accuracy. Yet, the best system comes from the highest dimensional input, where the \ac{RSSI} is combined with the \ac{IMU} and \ac{UWB} system; allowing even in 99\% of the cases to be closer than \SI{1}{\metre}. It becomes apparent that joining sensors even in this vanilla case allows for a better performance.
Note: This is on the same data-set thus it does not give an indication on the generalizability, which will be investigated next.

\subsection{Time stability (generalizability)}
\begin{figure}[H]
    \centering
    \includegraphics[]{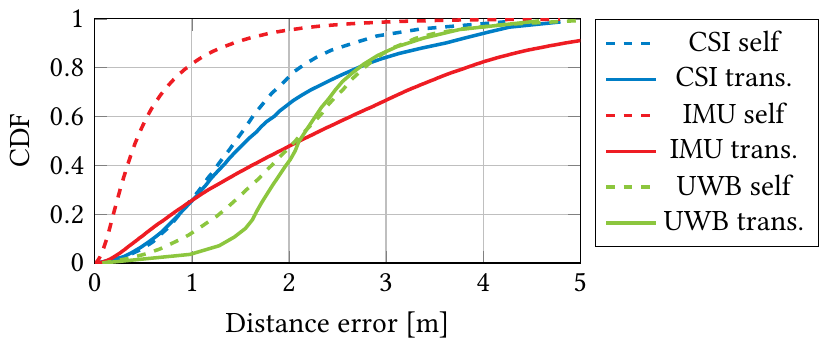}
    \vspace*{-0.4cm}
    \caption{Performance of the vanilla \ac{NN} from the morning to the evening data-set.} 
    \label{fig:PerfNNTranslation}
    \vspace{-0.4cm}
\end{figure}
Fig.~\ref{fig:PerfNNTranslation} demonstrates the performance of the \ac{NN} being trained on the Dataset 1 and evaluated on the Dataset 2. The solid curve gives the self performance meaning on the same 90/10 split dataset and the dashed curve gives the generalizability by evaluating without taking any data from Dataset 2 the performance. We show that the \ac{IMU} system performs very well being on exactly the same configuration, but small changes in the position or the rotation removes any advantage of the system. So a better generalization method needs to be applied to make \ac{IMU} data viable in itself. It can be seen that the \ac{UWB} system also did some form of over-fitting on the actual data as it has a better lower part of the \ac{CDF}, but yet it achieves similar performance on the overall data-set, showing that it generalized well. The best generalization is achieved by the \ac{CSI} approach were no over-fitting happened, but only a small degradation of the curve occurred. It shows that the measurement is reproducible and viable to do such kind of experiments. The results indicate that we can use a combination of sensor readings to achieve a reasonable performance level.

\begin{figure}[H]
\vspace*{-0.4cm}
    \centering
    \includegraphics[]{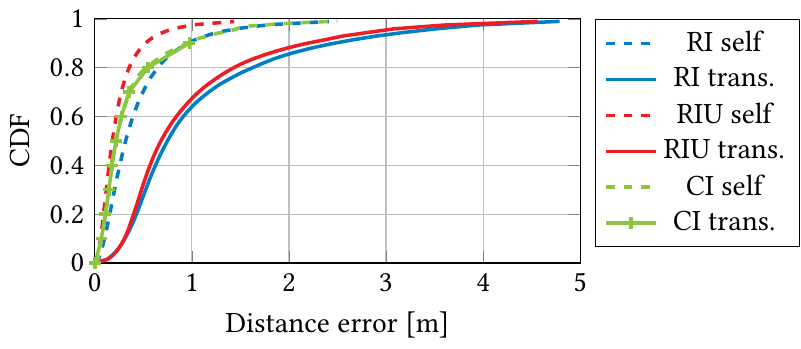}
    \vspace*{-0.4cm}
    \caption{Performance of the fused methods trying to predict from one data-set to another.} 
    \label{fig:FusedNNoverTime}
    \vspace{-0.4cm}
\end{figure}
Fig.~\ref{fig:FusedNNoverTime} depicts the performance of the fused techniques. All RSSI based techniques although having performed well on the first data-set, would require additional labels to stabilize performance. Thus, these methods require continuous updates to adjust to small changes in the scenario. In the case of the CSI-based method in combination with the IMU raw data it turns out that the curves exactly match, which was already hinted in the previous results, where the drift of the CSI was not as impacted as the other techniques. Thus the combination of the IMU and CSI data seem to be a solid combination for a reliable and robust positioning technique, achieving on average around \SI{10}{\centi\metre} of accuracy and in the 95\% around 1m accuracy.

\section{Conclusion}
In this paper we have presented a versatile and robust measurement system for creating large amount of labeled data for the investigation of localization techniques. We demonstrate that this dataset is viable for different fusion and mapping techniques. Different classical localization methods are compared with state-of-the-art \ac{NN} approaches demonstrating that the \ac{NN} techniques can handle outliers better than pre-defined algorithms. A vanilla approach to fuse the raw-data outputting the position is proposed. The \ac{CSI} data combined with the \ac{IMU} data have shown high performance, indicating the combination of those techniques to be a generalized high quality localization technique.

\vspace{-0.15cm}
\bibliographystyle{IEEEtran}

\bibliography{biblio}


\end{document}